\RequirePackage{fix-cm}
\documentclass[%
 reprint,
superscriptaddress,
 amsmath,amssymb,
 aps,
prl,
]{revtex4-2}

\usepackage{array}[=2016-10-06]

\usepackage{graphicx}
\usepackage{dcolumn}
\usepackage{bm}
\usepackage{color}
\usepackage{hyperref}

\makeatletter
\AtBeginDocument{\let\NAT@sort\z@\let\NAT@cmprs\@ne}
\makeatother

\usepackage{mathtools,booktabs,array,enumitem,pgfplots}
\usepgfplotslibrary{groupplots}
\pgfplotsset{compat=1.18}
\hypersetup{hidelinks,pdftitle={Efficient Quantum Optimization via Dissipative False Vacuum Decay: Main Text and Supplementary Material},pdfauthor={Zhi-Hao Huang, Tian-Mu Hu, Shuaifan Cao, Ye Lu, Xiaopeng Li}}

\makeatletter
\newcommand{\SMsection}[1]{%
  \par\addvspace{3.5ex plus 1ex minus .2ex}%
  \refstepcounter{section}\setcounter{subsection}{0}%
  \addcontentsline{smtoc}{section}{\protect\numberline{\thesection}#1}%
  \noindent{\Large\bfseries\thesection\quad #1\par}%
  \nobreak\vspace{2.3ex plus .2ex}\@afterindentfalse\@afterheading}
\newcommand{\SMsubsection}[1]{%
  \par\addvspace{3.25ex plus 1ex minus .2ex}%
  \refstepcounter{subsection}%
  \addcontentsline{smtoc}{subsection}{\protect\numberline{\thesubsection}#1}%
  \noindent{\large\bfseries\thesubsection\quad #1\par}%
  \nobreak\vspace{1.5ex plus .2ex}\@afterindentfalse\@afterheading}
\newcommand{\SMparagraph}[1]{\par\addvspace{1.3ex}\noindent\textbf{#1}\quad\ignorespaces}
\newcommand{\SMtableofcontents}{%
  \par\addvspace{2ex}\noindent{\Large\bfseries Contents\par}\vspace{1ex}%
  \begingroup
  \def\l@section##1##2{\addpenalty{-\@highpenalty}\vskip .65em%
    \@dottedtocline{1}{0em}{1.8em}{\bfseries ##1}{\bfseries ##2}}%
  \def\l@subsection{\@dottedtocline{2}{1.8em}{2.8em}}%
  \@starttoc{smtoc}\endgroup}
\newcounter{smref}
\newcommand{\SMcite}[1]{[%
  \begingroup\def\SMsep{}%
  \@for\SMkey:=#1\do{\SMsep\ref{smref:\SMkey}\def\SMsep{,\,}}%
  \endgroup]}
\newcommand{\SMbibitem}[1]{\item\label{smref:#1}}
\newenvironment{SMbibliography}{%
  \par\addvspace{3.5ex}\noindent{\Large\bfseries References\par}\vspace{1.5ex}%
  \small\begin{list}{[\arabic{smref}]}{%
    \usecounter{smref}\settowidth{\labelwidth}{[99]}%
    \setlength{\leftmargin}{\labelwidth}\addtolength{\leftmargin}{\labelsep}%
    \setlength{\itemsep}{.3em}\setlength{\parsep}{0pt}}%
}{\end{list}}
\makeatother

\begin{document}

\preprint{APS/123-QED}


\title{Efficient Quantum Optimization via Dissipative False Vacuum Decay}
\author{Zhi-Hao Huang}
\thanks{These authors contribute equally to this work}
\affiliation{State Key Laboratory of Surface Physics, Key Laboratory of Micro and Nano Photonic Structures (MOE), and Department of Physics, Fudan University, Shanghai 200433, China }
\affiliation{Shanghai Branch, Hefei National Laboratory, Shanghai 201315, China}
\affiliation{Shanghai Qi Zhi Institute, AI Tower, Xuhui District, Shanghai 200232, China}

\author{Tian-Mu Hu}
\thanks{These authors contribute equally to this work}
\affiliation{State Key Laboratory of Surface Physics, Key Laboratory of Micro and Nano Photonic Structures (MOE), and Department of Physics, Fudan University, Shanghai 200433, China }
\affiliation{Shanghai Branch, Hefei National Laboratory, Shanghai 201315, China}
\affiliation{Shanghai Qi Zhi Institute, AI Tower, Xuhui District, Shanghai 200232, China}
\affiliation{College of integrated Circuits and Micro-nano electronics, Fudan University, Shanghai 200433, China}

\author{Shuaifan Cao}
\thanks{These authors contribute equally to this work}
\affiliation{State Key Laboratory of Surface Physics, Key Laboratory of Micro and Nano Photonic Structures (MOE), and Department of Physics, Fudan University, Shanghai 200433, China }
\affiliation{Shanghai Branch, Hefei National Laboratory, Shanghai 201315, China}
\affiliation{Shanghai Qi Zhi Institute, AI Tower, Xuhui District, Shanghai 200232, China}

\author{Ye Lu}%
\affiliation{College of integrated Circuits and Micro-nano electronics, Fudan University, Shanghai 200433, China}

\author{Xiaopeng Li}%
  \email{xiaopeng\underline{ }li@fudan.edu.cn}
\affiliation{State Key Laboratory of Surface Physics, Key Laboratory of Micro and Nano Photonic Structures (MOE), and Department of Physics, Fudan University, Shanghai 200433, China }
\affiliation{Shanghai Branch, Hefei National Laboratory, Shanghai 201315, China}
\affiliation{Shanghai Qi Zhi Institute, AI Tower, Xuhui District, Shanghai 200232, China}
\affiliation{Buchou Quantum Technology Co., Ltd, Shanghai, China}
%





\begin{abstract}
Efficient quantum solvers for computationally hard problems
are highly desirable for near-term quantum platforms.
Here, we propose a cavity-based dissipative quantum solver
for a class of NP-hard combinatorial optimization problems
encoded in atomic-spin Hamiltonians with a few discrete single-spin-flip transition energies. We show that the cavity-based quantum solver exploits dissipative
false-vacuum decay in configuration space to facilitate escape from metastable local minima in rugged energy landscapes. Numerical studies of MaxCut and Maximum Independent Set identify favorable parameter regions for ground-state preparation. Based on the finite-size scalings of the median mixing time and instance-averaged ground-state preparation probability, we estimate time complexities for our solver that are more favorable than those of standard Grover search. These results highlight cavity dissipation as a controllable resource for quantum optimization.
\end{abstract}


\maketitle

%
\paragraph*{Introduction.}
False-vacuum decay is a paradigmatic manifestation of metastability that arises across a wide range of physical scales~\cite{Coleman1977,Fialko2015,Zenesini2024}. In this process, a long-lived many-body state escapes from a local energy minimum via quantum tunneling or thermal activation and evolves toward a more stable, lower-energy phase, typically through the nucleation and growth of bubbles or domains~\cite{CallanColeman1977,Linde1981}. More broadly, analogous transitions between metastable states provide a useful framework for understanding tipping phenomena in complex systems, including the possible transition of the Atlantic Meridional Overturning Circulation to a weakened or collapsed state, with potentially profound consequences for global environmental change \cite{Hawkins2011,vanWesten2024}. Closely related nucleation dynamics also occur in quantum many-body systems, including spin models and atomic superfluids \cite{Fialko2015,Lagnese2021,Darbha2024}. Recent advances in quantum simulation and quantum computing have enabled controlled experimental studies of false-vacuum decay and its associated real-time dynamics \cite{Zenesini2024,Vodeb2025,Chao2026}. 

In quantum computing, combinatorial optimization has attracted considerable interest because improved solutions to computationally challenging problems could provide an important route toward practical quantum advantage \cite{Farhi2001,PerdomoOrtiz2012,Grass2016,Hastings2021,Ebadi2022}. Two prominent paradigms are circuit-based variational algorithms~\cite{Peruzzo2014,CerezoVQA2021}, such as the Quantum Approximate Optimization Algorithm (QAOA) \cite{Farhi2014,Wecker2016,Zhou2020,Luo2025}, and Hamiltonian-based approaches, such as quantum annealing \cite{Kadowaki1998,Farhi2000,Das2008,Boixo2016,Denchev2016,Albash2018}. Their performance, however, is often limited by complex optimization landscapes and trapping in local minima. In QAOA, the classical optimization of variational parameters can stagnate in suboptimal regions~\cite{Zhou2020,Luo2025} and may be further hindered by barren plateaus \cite{McClean2018,Cerezo2021}, where gradients become exponentially suppressed. In quantum annealing, small spectral gaps and metastable configurations can induce diabatic transitions~\cite{Young2010,Jorg2010,Altshuler2010} or exponentially slow relaxation, preventing the system from efficiently reaching the ground state~\cite{Boixo2016,Kechedzhi2016}. Overcoming such metastability-induced bottlenecks is therefore central to developing efficient quantum optimization algorithms~\cite{Farhi2002,Wang2022}.

In this Letter, we introduce a cavity-based quantum solver that encodes combinatorial optimization problems into spin Hamiltonians~\cite{Ye2023,Ye2024} and exploits dissipative false-vacuum decay to overcome metastable trapping in rugged energy landscapes. By coupling a problem Hamiltonian with some discrete single-spin-flip transition energies to a tunable cavity mode, we engineer an interplay between coherent quantum tunneling and dissipative energy relaxation. Specifically, (i) the cavity mode facilitates resonant and near-resonant tunneling transitions that enable metastable spin configurations to escape false vacua, while (ii) cavity photon loss provides an irreversible relaxation channel and favors population transfer toward lower-energy configurations. Together, these processes establish systematic decay pathways from metastable states toward the ground-state manifold. We demonstrate this mechanism for the MaxCut and Maximum Independent Set (MIS) problems and find favorable finite-size scaling of the solution efficiency. Our results establish cavity dissipation as a controllable computational resource for quantum optimization, providing an efficient route to escape from local vacuum trapping in challenging combinatorial problems.
\begin{figure*}[htb]
    \centering
    \includegraphics[scale=1.1]{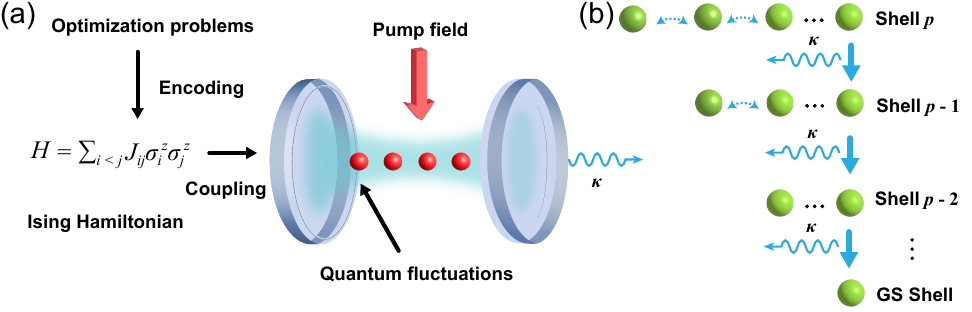}
    \caption{
    Schematic illustration of the cavity-based dissipative quantum solver and its many-body relaxation mechanism. (a) Physical realization. A combinatorial optimization problem is encoded in an atomic-spin Hamiltonian $H_{\mathrm{problem}}$, whose ground-state manifold represents the optimal solutions. The spin system is collectively coupled to a lossy single-mode optical cavity. Cavity-field quantum fluctuations mediate coherent spin-photon transitions through the atom-photon interaction, while the effective cavity-mode energy $\Delta$ in the rotating frame can be tuned by the pump field. Photon leakage at rate $\kappa$ provides an irreversible dissipation channel.
    (b) Energy-shell picture of the dissipative dynamics in the exactly resonant limit. When $\Delta$ matches a characteristic transition energy of $H_{\mathrm{problem}}$, composite spin--photon states $|E_{\alpha_q}\rangle\otimes|q\rangle$ satisfying $E_{\alpha_q}+q\Delta=\mathcal{E}$ become degenerate and form an energy shell with total energy $\mathcal{E}$. The green spheres denote these composite states. Provided that the corresponding spin-flip matrix elements are nonzero, the cavity-induced interaction coherently hybridizes allowed states within each shell, as indicated by the dashed arrows. Photon loss induces transitions $|E_{\alpha_q}\rangle\otimes|q\rangle\rightarrow|E_{\alpha_q}\rangle\otimes|q-1\rangle$, transferring population to a neighboring shell with lower total energy. The repeated action of coherent intra-shell mixing and dissipative inter-shell relaxation generates a net population flow toward the ground-state manifold of $H_{\mathrm{problem}}$.
    }
    \label{fig:fig1}
\end{figure*}
\paragraph*{The dissipative quantum model.—}
We first introduce the physical realization of our cavity-based quantum solver. 
Many combinatorial optimization problems can be encoded into atomic spin Hamiltonians~\cite{Lucas2014,Qiu2020} with discrete single-spin-flip transition energies. As a representative example, we consider the Ising-type Hamiltonian
$H_{\mathrm{problem}}=\sum_{i<j}J_{ij}\sigma_i^z\sigma_j^z$,
where $\sigma_i^z$ denotes the Pauli-$z$ operator at site $i$, and $J_{ij}$ represents the graph-dependent coupling. For an unweighted graph, $J_{ij}=J A_{ij}$, where $A_{ij}$ is the adjacency matrix.
As illustrated in Fig.~\ref{fig:fig1}(a), we couple the spin system to a lossy single-mode optical cavity, where cavity field quantum fluctuations provide a coherent spin-photon energy exchange channel. The resulting atom-photon interaction is described by
$H_{\mathrm{coupling}}=\lambda(a^\dagger+a)\sum_i\sigma_i^x$,
where $a^\dagger$ and $a$ are the photon creation and annihilation operators, respectively, and $\lambda$ denotes the atom-photon coupling strength.  The cavity mode is described by
$H_{\mathrm{photon}}=\Delta a^\dagger a$,
where $\Delta$ represents the effective cavity mode detuning in the rotating frame~\cite{Dimer2007,Baumann2010,Ritsch2013}. By tuning the pump frequency, $\Delta$ can be adjusted to control the resonance between the cavity field and characteristic transition energies of $H_{\mathrm{problem}}$. 
The spin states are labeled according to their problem Hamiltonian eigenvalues as $|E_m\rangle$. The photon states are denoted as $|q\rangle$, with $q$ the photon number.
The coupled spin-photon system is initialized in an equal-weight superposition of the eigenstates of $H_{\mathrm{problem}}$ together with the cavity vacuum,
$|\psi(0)\rangle\propto\sum_m|E_m\rangle\otimes|q=0\rangle$.
The subsequent dissipative evolution is governed by the Lindblad master equation~\cite{Gorini1976,Lindblad1976}
\begin{equation}
\frac{d\rho}{dt}
=
-i[H_{\mathrm{total}},\rho]
+
\kappa
\left(
a\rho a^\dagger
-\frac{1}{2}a^\dagger a\rho
-\frac{1}{2}\rho a^\dagger a
\right),\label{1}
\end{equation}
where
$H_{\mathrm{total}}=H_{\mathrm{problem}}+H_{\mathrm{coupling}}+H_{\mathrm{photon}}$.
 Here, $\rho$ denotes the density matrix of the coupled spin-photon system, and the Lindblad jump operator $a$ describes photon leakage from the cavity into the environment.

\paragraph*{Dissipative relaxation mechanism.—}
The dissipative dynamics of our cavity-based quantum solver is governed by  an interplay between coherent quantum tunneling among spin configurations and dissipative energy relaxation. The cavity-mode energy $\Delta$ sets the detunings of relevant coherent spin-photon transitions, while $\lambda$ sets their coupling matrix elements. The photon-loss rate $\kappa$ introduces a finite dissipative linewidth and renders the dynamics irreversible. Two spin configurations $|E_m\rangle$ and $|E_n\rangle$ that can be connected by a single spin flip, i.e.,  $\langle E_m|\sum_i\sigma_i^x|E_n\rangle\neq0$, are hereafter referred to as {\it nearby spin states}. 


We consider combinatorial optimization problems whose problem Hamiltonians have integer-valued spectra in the chosen energy unit. Examples include SAT, MaxCut, and vertex-cover \cite{Lucas2014}. To illustrate the relaxation mechanism, we first consider the exactly resonant limit, where the difference between the integer-valued energies of the spin states can be resonantly compensated by the photon energy. Composite states $|E_m\rangle\otimes|q\rangle$ satisfying
$E_m+q\Delta=\mathcal E$ form an energy shell with bare total energy $\mathcal E$, as shown in Fig.~\ref{fig:fig1}. The states within each energy shell are resonantly coupled by the cavity-induced interactions, so they can propagate freely in that shell, and the corresponding quantum dynamics are ballistic, in the absence of localization~\cite{Childs2002QuantumWalk,Perets2008QuantumWalk}. The photon leakage drives dissipative dynamics from 
$|E_m\rangle\otimes|q\rangle$ 
to
$|E_m\rangle\otimes|q-1\rangle$. 
In this process, the total energy is lowered by $\Delta$, causing an irreversible energy flow. 
Although the spin states remain the same in the decay process, the phase space shrinks for the maximal spin-energy accessible becomes smaller.

The exactly resonant energy-shell picture represents the simplest limit of this dynamics. For finite coupling and cavity linewidth, near-resonant transitions remain active even when the relevant composite states are not exactly degenerate. Here, near-resonant transitions have a small detuning from exact resonance but remain dynamically active at the couplings and linewidths considered.
Consequently, the optimal $\Delta$ need not coincide with any isolated transition energy.

Together, the intra-shell ballistic dynamics and dissipative
inter-shell relaxation favor a net population flow toward
lower-energy configurations of $H_{\mathrm{problem}}$, and the
dynamics eventually approach a steady state of the master
equation~\eqref{1}. In the weak-coupling limit, the steady-state
density matrix takes the form
\begin{equation}
\begin{aligned}
\rho_{\mathrm{ss}}
&=
\rho_{\mathrm{s}}^{(0)}
\otimes|0\rangle\langle0|
+\mathcal{O}(\lambda),
\qquad
\mathcal{K}_2
\!\left[\rho_{\mathrm{s}}^{(0)}\right]=0 .
\end{aligned}
\label{eq:steadystate}
\end{equation}
Here, $\rho_{\rm s}^{(0)}$ is a positive semidefinite,
normalized spin density operator satisfying
$[H_{\rm problem},\rho_{\rm s}^{(0)}]=0$.
The operator $\mathcal K_2$ is the second-order effective
generator acting in the unperturbed stationary
subspace, with the overall factor $\lambda^2$ removed \cite{Kessler2012,SM}.
The zeroth-order steady state provides a weak-coupling
reference for ground-state preparation \cite{SM}.

\begin{figure}[!t]
    \includegraphics[scale=0.21]{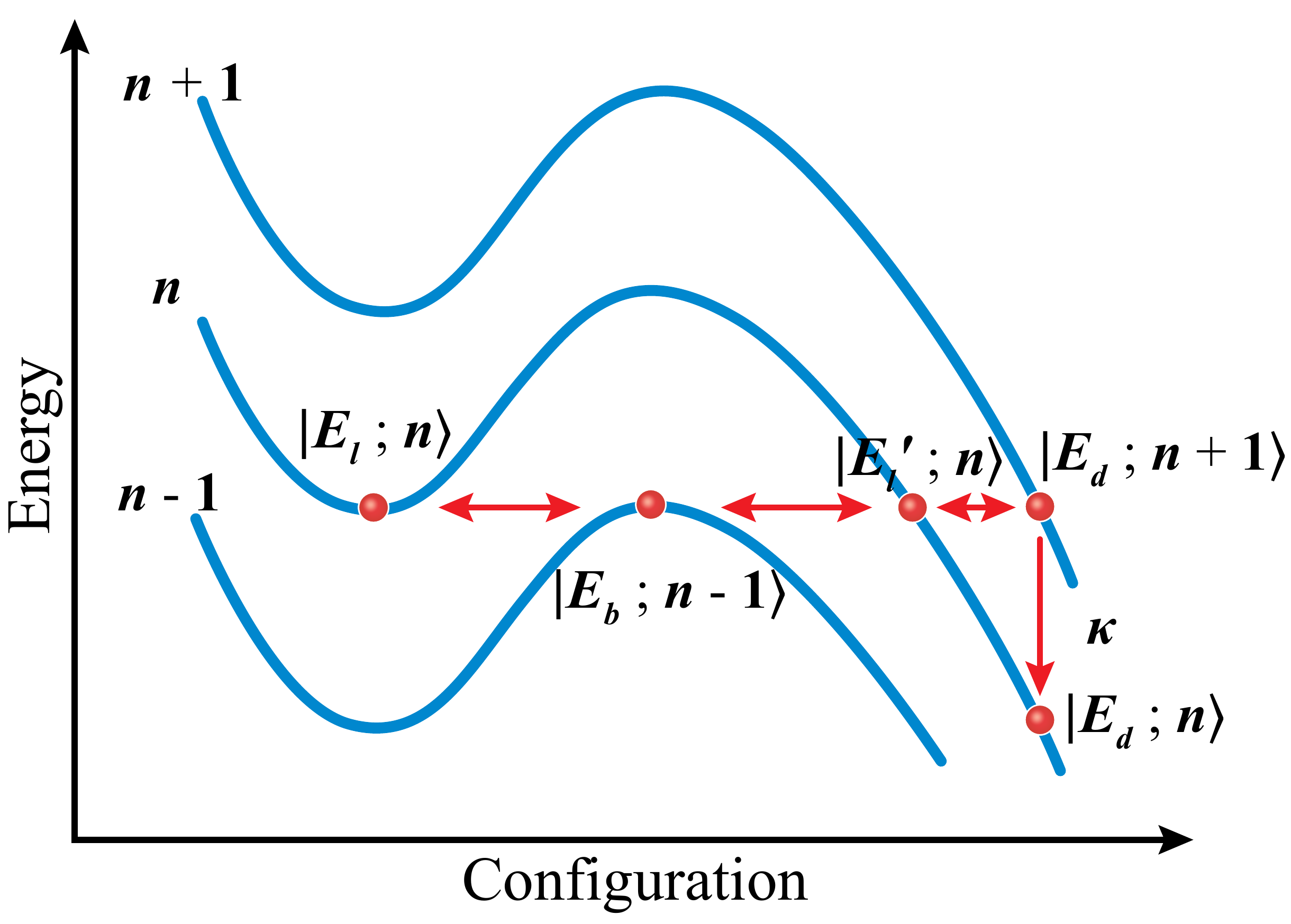}
\caption{
Dissipative false vacuum decay in configuration space.
The blue curves represent the effective composite energy landscapes
$\mathcal{E}_{q}(\mathcal{C})=E(\mathcal{C})+q\Delta$
in the neighboring cavity-photon sectors $q=n-1,n,$ and $n+1$, where $\mathcal{C}$ labels a spin configuration. Starting from a composite state $|E_\ell;n\rangle$ with
$n>0$ and a spin configuration at a nonglobal local minimum,
population can leave the original basin through the photon-assisted resonant pathway
$|E_\ell;n\rangle\leftrightarrow|E_b;n-1\rangle\leftrightarrow|E_\ell';n\rangle$.
Here, $|E_b\rangle$ is a higher-energy bridge configuration, while $|E_\ell'\rangle$ is a distinct degenerate configuration located outside the original local basin. When $E_b-E_\ell=\Delta$, these composite states are resonantly connected, as indicated by the horizontal double arrows. A subsequent downhill transition,
$|E_\ell';n\rangle\leftrightarrow|E_d;n+1\rangle$ with $E_\ell'-E_d=\Delta$,
converts the decrease in spin energy into a real cavity photon. Photon leakage,
$|E_d;n+1\rangle\xrightarrow{\kappa}|E_d;n\rangle$,
removes this photon, lowers the total spin-photon energy by $\Delta$, and suppresses coherent population return. The figure illustrates one representative pathway in the exactly resonant limit.}
\label{fig2}
\end{figure}
\paragraph*{Dissipative false-vacuum decay in optimization landscapes.—}

We now show how the above mechanism enables escape from
metastable states in rugged optimization landscapes.
A spin configuration at a local minimum plays
the role of a false vacuum, whereas the ground-state
manifold represents the true vacuum. Here, cavity-assisted
resonant and near-resonant transitions out of the local basin,
followed by photon-loss-induced relaxation toward lower-energy configurations, realize dissipative false-vacuum decay
in configuration space.

Conventional simulated annealing escapes local minima
through thermally activated barrier crossing on the bare
cost landscape~\cite{Kirkpatrick1983}, whereas quantum annealing and quantum walks,
in their standard coherent formulations, rely on tunneling
or coherent hopping and interference~\cite{Farhi1998,Kempe2003,Kadowaki1998,Albash2018} without introducing
auxiliary photon-number replicas of the landscape.
Our solver instead embeds the spin landscape into a ladder
of photon-dressed sectors,
$\mathcal{E}_q(\mathcal{C})=E(\mathcal{C})+q\Delta$,
where resonant and near-resonant transitions between
neighboring sectors provide additional escape pathways.
Although the cavity is initialized in the vacuum state
in our protocol, the many-body dynamics can transiently
populate finite-photon sectors. As shown in Fig.~2, we consider a representative composite state
$|E_\ell;n\rangle$ with $n>0$, whose spin configuration
lies at a local minimum. Here, $|E_\ell\rangle$ denotes a false-vacuum spin state: all nearby spin states obtained
by single-spin flips have higher energies.
The state $|E_\ell'\rangle$ denotes a distinct spin configuration
with $E_\ell'=E_\ell$ located outside the original local basin.
We further denote by $|E_b\rangle$ a higher-energy bridge
configuration and by $|E_d\rangle$ a lower-energy configuration.
The spin configurations connected by each coherent transition
below are assumed to differ by a single spin flip.
The relevant detunings are defined as
$\delta_r=E_b-E_ \ell-\Delta$ and $\delta_d=E_\ell'-E_d-\Delta$.

For $\delta_r$ or $\delta_d$ close to zero, the corresponding
composite states become degenerate or nearly degenerate in
bare total spin-photon energy, giving rise to resonant or
near-resonant transition pathways.
The system can leave the local basin through the
photon-assisted pathway
$|E_\ell;n\rangle
\leftrightarrow
|E_b;n-1\rangle
\leftrightarrow
|E_\ell';n\rangle$. Once population reaches $|E_\ell';n\rangle$, the downhill transition channel $|E_\ell';n\rangle\leftrightarrow|E_d;n+1\rangle$
transfers the energy to the cavity, after which
the additional photon can leak out through
$|E_d;n+1\rangle\xrightarrow{\kappa}|E_d;n\rangle$.
Photon leakage removes this additional photon and provides
an irreversible energy-loss channel, favoring net downhill
relaxation.
This two-stage pathway illustrates the dissipative false-vacuum decay in configuration space.

\begin{figure}[!t]
    \includegraphics[scale=1.4]{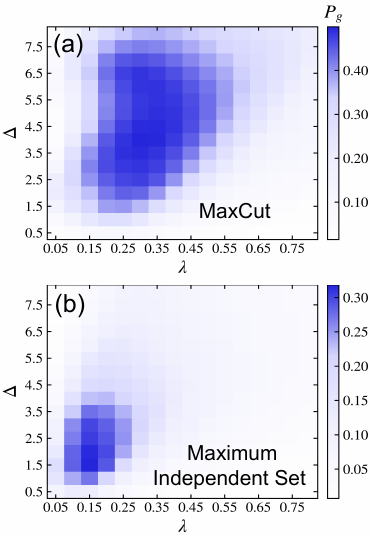}
    \caption{
    Instance-averaged ground-state preparation probability 
    $\overline{P_g} (T=200)$ 
    for Max-Cut and MIS on random 3-regular graphs of size $N=10$. At each parameter point $(\Delta,\lambda)$, $P_g$ is averaged over ten independently generated graph instances. For each problem, the same ten graph instances are used
    throughout the parameter scan. The atom-photon coupling is time independent throughout the evolution, $\lambda(t)=\lambda$. The photon-loss rate is fixed at $\kappa=1.0$, and the cavity Hilbert space is truncated to $d_{\mathrm{photon}}=4$. For Max-Cut, energies and rates are expressed in units of $J$. For MIS, energies and rates are expressed in units of $\delta$.
    (a) For Max-Cut with unit edge weights, regions of comparatively high ground-state preparation probability emerge around $\Delta/J\simeq3.0$--$3.5$ and $\Delta/J\simeq6.0$--$6.5$, with an optimal intermediate coupling near $\lambda/J\simeq0.25$--$0.3$.
    (b) For MIS, the optimal region is centered near $\Delta/\delta\simeq2.0$ and $\lambda/\delta\simeq0.15$.
    }
    \label{fig3}
\end{figure}
\paragraph*{Performance on NP-hard problems.—}
We test the cavity-based quantum solver on MaxCut and MIS
using random 3-regular graphs with $N=10$.
The initial spin-cavity state is
$|\psi(0)\rangle
=2^{-N/2}\sum_{m=0}^{2^N-1}|E_m\rangle\otimes|0\rangle$,
corresponding to an equal-weight superposition of all spin states and the cavity vacuum.
The coupled system evolves according to Eq.~\eqref{1}
up to $T=200$.
The success probability is defined as
$P_g(T)
=\operatorname{Tr}[(\Pi_g\otimes I_{\mathrm{cav}})\rho(T)]$,
where $\Pi_g$ projects onto the ground-state manifold
of the problem Hamiltonian.
The results in Fig.~3 are averaged over ten independently
generated graph instances.

For MaxCut with unit edge weights, the problem Hamiltonian is $H_{\mathrm{MaxCut}}
=J\sum_{(ij)\in\mathcal{E}(\mathcal{G})}
\sigma_i^z\sigma_j^z$~\cite{Farhi2014},
with $J=1$.
With the spin eigenvalues
$s_i=\pm1$, flipping spin $i$ produces an energy change of magnitude
$|\Delta E|
=|-2Js_i\sum_{j\in\partial i}s_j|
\in\{2,6\}$.
For MIS, we consider
$H_{\mathrm{MIS}}
=-\delta\sum_i n_i
+J\sum_{(ij)\in\mathcal{E}(\mathcal{G})}n_i n_j$,
where $\delta=1$, $J=2$, and $n_i=|1\rangle_i\langle1|$.
The first term favors occupied vertices, whereas the second
penalizes adjacent occupations, so that the ground-state
    manifold encodes the maximum independent sets~\cite{Ebadi2022,Pichler2018,Nguyen2023}.
Changing the occupation of a vertex with
$k\in\{0,1,2,3\}$ occupied neighbors produces an energy
change of magnitude
$|\Delta E|=|-1+2k|\in\{1,3,5\}$.

As shown in Fig.~3, MaxCut exhibits regions of comparatively
high ground-state preparation probability around
$\Delta\simeq3.0$--$3.5$ and $6.0$--$6.5$,
with optimal coupling values $\lambda\simeq0.25$--$0.3$,
whereas the MIS optimum occurs near
$(\Delta,\lambda)=(2.0,0.15)$.
The higher-frequency MaxCut window lies close to the bare
single-spin-flip transition energy $|\Delta E|=6$.
Its width is consistent with the finite cavity linewidth
and coupling, while any shift from the bare resonance
may reflect cavity-induced level dressing and competition
between population transfer and coherent backflow.
Neither the lower-frequency MaxCut window nor the MIS
optimum coincides with a bare single-spin-flip transition energy, indicating that matching $\Delta$ to a single
bare transition energy does not fully explain optimal performance.
These results are consistent with the above mechanism
involving multiple resonant and near-resonant transitions whose relative strengths depend on $\Delta$.
Detuning from an individual bare transition may retain
useful escape and relaxation channels while reducing
coherent backflow.
The preference for intermediate coupling suggests
a balance between population-transfer efficiency and
energy selectivity.
Photon leakage provides an irreversible energy-loss
channel that supports relaxation toward the ground-state
manifold.


\begin{figure}[!t]
    \includegraphics[scale=0.38]{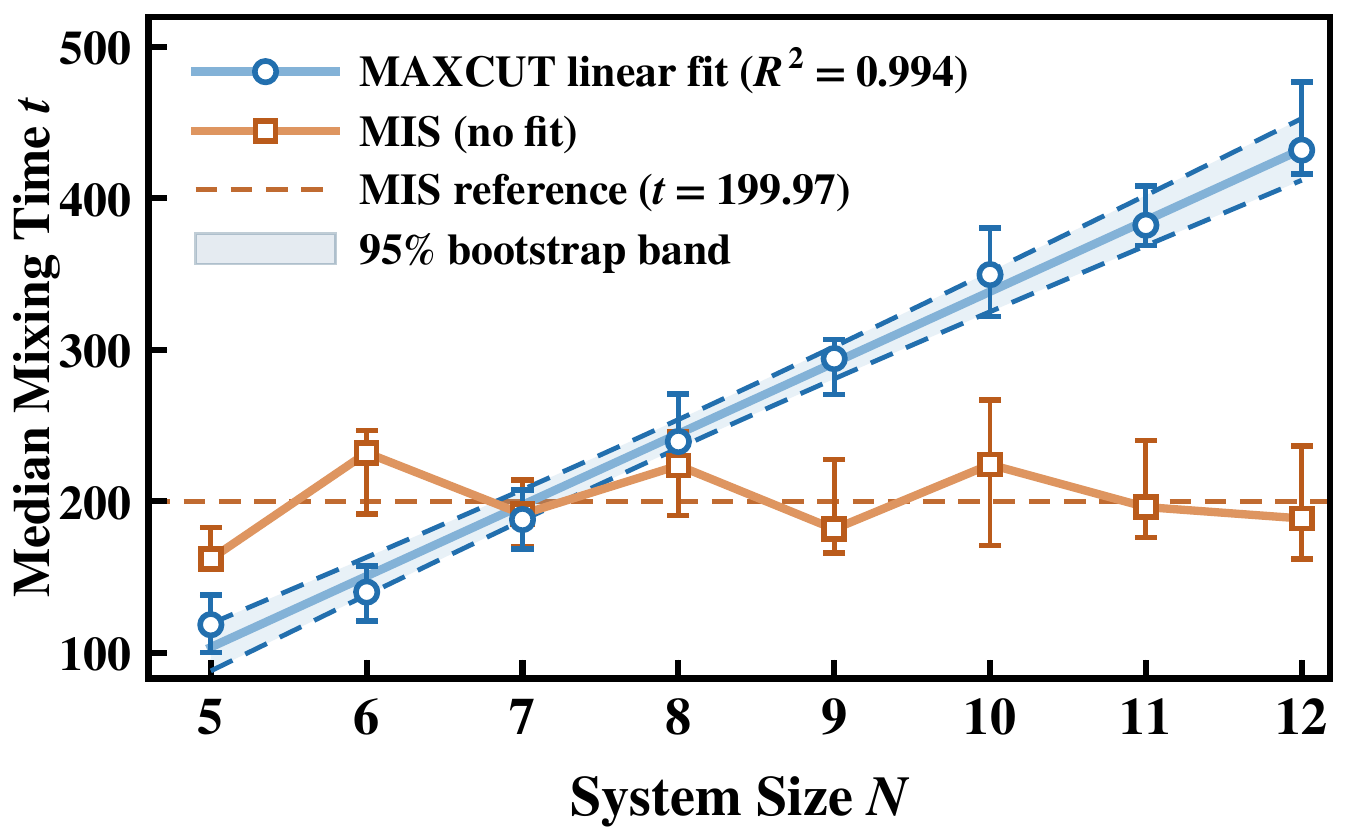}
    \caption{Finite-size mixing-time scaling of the cavity-based quantum solver for MaxCut (blue circles)
and MIS (orange squares) on random connected graphs with
ensemble-averaged degree $\langle k\rangle=3$.
Each point is the median over 100 instances per problem
at $N=5,\ldots,12$.
For each instance, the mixing time is defined
as the earliest time $T_0$ satisfying
$\max_{1\leq j\leq1000}
|P_g(T_0+j\delta t)-P_g(T_0)|/P_g(T_0)\leq0.01$,
where $\delta t=0.05$, corresponding to a subsequent
monitoring window of 50 time units.
Error bars indicate bootstrap uncertainties in the medians.
The blue solid line is a linear fit ($R^2=0.994$),
with shading showing its 95\% bootstrap confidence band.
The dashed line at $t=200$ is an MIS reference, not a fit.
Parameters:
$(\Delta/J,\lambda/J,\kappa/J)=(3.0,0.25,1.0)$ for MaxCut
and $(\Delta/\delta,\lambda/\delta,J/\delta,\kappa/\delta)
=(2.0,0.15,2.0,1.0)$ for MIS.
Both problems use a truncated cavity Hilbert-space dimension
$d_{\mathrm{photon}}=4$.
}
    \label{fig4}
\end{figure} 

    \paragraph*{Finite-size mixing-time scaling.—}
    Finally, we examine finite-size relaxation of our cavity-based quantum solver at $(\Delta,\lambda)=(3.0,0.25)$ for MaxCut and
    $(2.0,0.15)$ for MIS, selected from the preceding
    finite-time scans and fixed across all sizes.
    Figure~\ref{fig4} shows the median mixing time $t_{\rm mix}$ for $N=5$--$12$ on random connected graphs with
    ensemble-averaged degree $\overline{k}=3$.
    For each instance, the mixing time $T_0$ is operationally defined as the earliest time at which the ground-state preparation probability remains stable within a $1\%$ relative variation over the subsequent 50 time units. MaxCut exhibits approximately linear growth,
    $t_{\rm mix}\simeq a+bN$, whereas the MIS mixing times
    fluctuate around $200$ without a clear systematic increase.
    
    To estimate the time complexity of our solver, we combine the mixing-time scaling with the instance-averaged ground-state preparation probability. Averaging the probabilities at
    each instance's mixing time $T_0$ gives the approximate 
    success probability scaling $\overline{P_g}(T_0)\propto2^{-\alpha N}$,
    with $\alpha\simeq0.16$ for MaxCut and $0.25$ for MIS.
    These exponents are smaller than the corresponding weak-coupling reference values $\alpha_0\simeq0.19$ and $0.29$, fitted to the zeroth-order steady-state results of Eq.~\eqref{eq:steadystate}\cite{SM}, indicating improved ground-state preparation at finite coupling. Accounting for repeated preparation at a fixed target success probability, we obtain time complexities of $\mathcal O(N2^{0.16N})$ for MaxCut and $\mathcal O(2^{0.25N})$ for MIS \cite{SM}. For comparison, Grover search requires $\mathcal O(2^{0.5N})$ oracle queries~\cite{Grover1997,Bennett1997,Boyer1998}. Thus, the estimated time complexities of our solver are more favorable than the scaling of Grover search. We expect the success probability scaling of our algorithm can be further improved by considering more complex dissipative quantum dynamics.

\paragraph*{Discussion and conclusion.—}
We have presented a cavity-based quantum solver combining
resonant and near-resonant coherent transitions with
photon-loss-induced relaxation. Their interplay underlies
the proposed mechanism of dissipative false-vacuum decay
in configuration space. For MaxCut and MIS, finite coupling
yields more favorable ground-state preparation scaling
than the zeroth-order steady-state reference.
By combining the median mixing times with the
instance-averaged ground-state preparation probabilities,
we obtain estimated time complexities of
$\mathcal{O}(N2^{0.16N})$ for MaxCut and
$\mathcal{O}(2^{0.25N})$ for MIS, which are more favorable
than the $\mathcal{O}(2^{0.5N})$ scaling of standard
Grover search.

This approach could also 
be integrated with quantum annealing,
where finite-rate passages through exponentially small
gaps can generate diabatic excitations in some optimization
problems. A subsequent cavity-cooling stage could help
remove the remaining excitation energy: spin-photon
coupling transfers energy released by transitions to
lower-energy spin configurations into cavity photons,
which then leak into the environment~\cite{Cho2015,Cho2024Cooling}.
Such a hybrid protocol could improve ground-state
preparation by combining coherent annealing with
dissipative relaxation~\cite{Amin2008,Albash2015}, rather than relying on adiabatic
evolution alone.
Extending the solver to a multimode cavity~\cite{Marsh2025} could allow
modes with different effective detunings to address distinct
spin-flip transition energies simultaneously.
This spectral flexibility could provide additional escape
and relaxation pathways.
Thus, a dissipation channel conventionally regarded as
detrimental to coherent quantum computation becomes
a computational resource for quantum optimization~\cite{Verstraete2009,Ding2024,Zhan2026}.

\paragraph*{Acknowledgments.}
We thank Xinxin Yang, Dong Chen, Kuangjie Chen for their valuable
discussions. This work was supported by the National
Basic Research Program of China (Grant No.~2021YFA1400900), and the Innovation Program
for Quantum Science and Technology of China (Grant No.~2024ZD0300100), and Shanghai
Municipal Science and Technology (Grant Nos.~26LZ0500100, 25TQ003, and 24DP2600100).

\bibliography{solverI}

\clearpage
\onecolumngrid
\clearpage
\begingroup
\makeatletter
\def\p@section{}\def\p@subsection{}\def\p@subsubsection{}
\renewcommand{\normalsize}{\@setfontsize\normalsize{11}{13.6}%
  \abovedisplayskip 11pt plus 3pt minus 6pt%
  \abovedisplayshortskip 0pt plus 3pt%
  \belowdisplayshortskip 6.5pt plus 3.5pt minus 3pt%
  \belowdisplayskip\abovedisplayskip}
\renewcommand{\small}{\@setfontsize\small{10}{12}}
\renewcommand{\footnotesize}{\@setfontsize\footnotesize{9}{11}}
\makeatother
\normalsize
\setlength{\parindent}{1.1em}
\setlength{\parskip}{0.12em}
\setlength{\emergencystretch}{2em}
\setlist{itemsep=3pt,topsep=5pt}
\allowdisplaybreaks[2]
\setcounter{section}{0}
\setcounter{subsection}{0}
\setcounter{equation}{0}
\setcounter{figure}{0}
\setcounter{table}{0}
\setcounter{secnumdepth}{2}
\setcounter{tocdepth}{2}
\renewcommand{\thesection}{\Alph{section}}
\renewcommand{\thesubsection}{\thesection.\arabic{subsection}}
\renewcommand{\theequation}{S\arabic{equation}}
\renewcommand{\thefigure}{S\arabic{figure}}
\renewcommand{\thetable}{S\arabic{table}}
\renewcommand{\theHsection}{SM.\arabic{section}}
\renewcommand{\theHsubsection}{SM.\arabic{section}.\arabic{subsection}}
\renewcommand{\theHequation}{SM.\arabic{equation}}
\renewcommand{\theHfigure}{SM.\arabic{figure}}
\renewcommand{\theHtable}{SM.\arabic{table}}
\phantomsection
\pdfbookmark[0]{Supplementary Material}{supplementary-material}

\begin{center}
{\large\bfseries Supplementary Material for}\par\vspace{0.35em}
{\Large\bfseries Efficient Quantum Optimization via\par Dissipative False Vacuum Decay}\par\vspace{0.8em}
{Zhi-Hao Huang\textsuperscript{1,2,3,*}, Tian-Mu Hu\textsuperscript{1,2,3,4,*}, Shuaifan Cao\textsuperscript{1,2,3,*},\par
Ye Lu\textsuperscript{4}, and Xiaopeng Li\textsuperscript{1,2,3,5,$\dagger$}}\par\vspace{0.65em}
{\small
\textsuperscript{1}State Key Laboratory of Surface Physics, Key Laboratory of Micro and Nano Photonic\par
Structures (MOE), and Department of Physics, Fudan University, Shanghai 200433, China\par
\textsuperscript{2}Shanghai Branch, Hefei National Laboratory, Shanghai 201315, China\par
\textsuperscript{3}Shanghai Qi Zhi Institute, AI Tower, Xuhui District, Shanghai 200232, China\par
\textsuperscript{4}College of Integrated Circuits and Micro-Nano Electronics,\par
Fudan University, Shanghai 200433, China\par
\textsuperscript{5}Buchou Quantum Technology Co., Ltd., Shanghai, China\par}
\end{center}
\begin{center}\footnotesize
$*$ These authors contribute equally to this work.\qquad
$\dagger$ \href{mailto:xiaopeng_li@fudan.edu.cn}{xiaopeng\_li@fudan.edu.cn}
\end{center}
\vspace{0.2em}
\noindent
This Supplementary Material provides the derivations and numerical methods supporting the main text. Section~\ref{sec:sm-model} derives the effective spin-photon coupling and specifies the graph ensembles and simulation settings. Section~\ref{sec:transition-network} describes allowed transitions and possible escape pathways in the configuration-space picture of false-vacuum decay. Section~\ref{sec:weak-coupling-steady-state} derives the second-order spin generator and its stationary-state condition, then presents quantum-trajectory simulations and finite-size probability fits. Section~\ref{sec:sm-numerics} explains the full-solver integration method and photon-cutoff comparison, defines mixing time and success probability, and gives a qualitative estimate of time complexity.
\vspace{0.3em}

\setcounter{tocdepth}{2}
\SMtableofcontents
\clearpage

\SMsection{Microscopic coupling and parameter settings}
\label{sec:sm-model}

\SMsubsection{Microscopic derivation of the spin--photon coupling}
\label{sec:sm-coupling}
We derive the coupling using two cavity-assisted Raman channels~\SMcite{Dimer2007}, with $\hbar=1$. Each atom has two long-lived states $|0\rangle_i,|1\rangle_i$ and two auxiliary excited states $|r\rangle_i,|s\rangle_i$. We write $P_{\mu i}=|\mu\rangle_i\langle\mu|$, $\sigma_i^+=|1\rangle_i\langle0|$, $\sigma_i^-=(\sigma_i^+)^\dagger$, and $n=a^\dagger a$. The diagonal operator $H_{\mathrm{enc}}$ describes the separately engineered graph interactions and local energies. We assume uniform optical couplings across the atoms.

\SMparagraph{Optical interaction and rotating-wave approximation.}
The atom--field interaction is
\[
 H_{\mathrm{dip}}=-\sum_i\boldsymbol d_i\cdot
 [\boldsymbol E_{\mathrm{cav}}(\boldsymbol r_i)+\boldsymbol E_{\mathrm L}(\boldsymbol r_i,t)].
\]
The cavity couples $|0\rangle\leftrightarrow|r\rangle$ and $|1\rangle\leftrightarrow|s\rangle$, while two lasers couple $|1\rangle\leftrightarrow|r\rangle$ and $|0\rangle\leftrightarrow|s\rangle$. Their frequencies and polarizations select these transitions. Projecting onto the four levels and applying the optical rotating-wave approximation gives
\begin{align}
 H_{\mathrm{lab}}(t)&=H_{\mathrm{enc}}+\omega_c n
 +\sum_i(\omega_1P_{1i}+\omega_rP_{ri}+\omega_sP_{si})+V_{\mathrm{lab}}(t),
 \label{eq:micro-lab}\\
 V_{\mathrm{lab}}(t)&=\sum_i\Big[g_ra|r\rangle_i\langle0|+g_sa|s\rangle_i\langle1|
 +\frac{\Omega_r}{2}e^{-i\omega_{Lr}t}|r\rangle_i\langle1|
 \nonumber\\[-2pt]
 &\hspace{3em}+\frac{\Omega_s}{2}e^{-i\omega_{Ls}t}|s\rangle_i\langle0|+\mathrm{H.c.}\Big].
 \label{eq:micro-lab-interaction}
\end{align}
Here $g_{r,s}$ are cavity coupling amplitudes and $\Omega_{r,s}$ are laser Rabi frequencies; their phases are included in these complex coefficients. The rotating-wave approximation drops terms oscillating at sums of optical frequencies, which are much faster than the couplings and detunings retained below.

\SMparagraph{Rotating frame.}
Define the reference frequencies and the rotation generator by
\begin{equation}
 \begin{aligned}
 \omega_{\mathrm{ref}}&=\frac{\omega_{Ls}-\omega_{Lr}}{2},\qquad
 \nu=\frac{\omega_{Ls}+\omega_{Lr}}{2},\\
 G&=\nu n+\sum_i\big(\nu P_{ri}+\omega_{Ls}P_{si}+\omega_{\mathrm{ref}}P_{1i}\big).
 \end{aligned}
 \label{eq:micro-frame-generator}
\end{equation}
With $U=e^{iGt}$, the rotating-frame Hamiltonian $H_R=UH_{\mathrm{lab}}U^\dagger-G$ is time independent:
\begin{equation}
 H_R=H_{\mathrm{enc}}+\Delta_0n+\epsilon\sum_iP_{1i}
 +\sum_i(\delta_rP_{ri}+\delta_sP_{si})+V_{\mathrm{opt}},
 \label{eq:micro-rotated-H}
\end{equation}
where $V_{\mathrm{opt}}$ is Eq.~\eqref{eq:micro-lab-interaction} with the laser phase factors removed, and
\begin{equation}
 \Delta_0=\omega_c-\nu,\qquad \epsilon=\omega_1-\omega_{\mathrm{ref}},\qquad
 \delta_r=\omega_r-\nu,\qquad \delta_s=\omega_s-\omega_{Ls}.
 \label{eq:micro-detunings}
\end{equation}
Thus the laser-frequency difference sets the spin reference frequency, while their average sets the cavity reference frequency.

\SMparagraph{Elimination of the auxiliary excited states.}
We take $|\delta_{r,s}|$ much larger than the optical coupling matrix elements for the occupied states, including photon-number and collective enhancement. They also exceed $|\Delta_0|$, $|\epsilon|$, $\kappa$, and the encoding-induced changes in the virtual-transition energies. Starting from the long-lived manifold, the atoms remain mainly in $|0\rangle$ and $|1\rangle$, with small auxiliary-state amplitudes that can be eliminated~\SMcite{ReiterSorensen2012}.

For clarity, first consider one atom and omit its label $i$. Write the atom--cavity state as $|\Psi\rangle=\sum_{\mu=0,1,r,s}|\mu\rangle|\psi_\mu\rangle$, where $|\psi_\mu\rangle$ is the unnormalized cavity state associated with atomic level $|\mu\rangle$. Projecting the Schr\"odinger equation $i\partial_t|\Psi\rangle=H_R|\Psi\rangle$ onto $\langle r|$ and $\langle s|$, and neglecting slow energy corrections compared with $\delta_{r,s}$, gives
\begin{equation}
 i\partial_t|\psi_r\rangle\simeq\delta_r|\psi_r\rangle+g_ra|\psi_0\rangle+\frac{\Omega_r}{2}|\psi_1\rangle,\qquad
 i\partial_t|\psi_s\rangle\simeq\delta_s|\psi_s\rangle+g_sa|\psi_1\rangle+\frac{\Omega_s}{2}|\psi_0\rangle.
 \label{eq:micro-auxiliary-amplitudes}
\end{equation}
Each equation contains the auxiliary-level detuning and its two optical couplings. To obtain the slow dynamics at large detuning, we neglect the rapid oscillations and set the auxiliary-amplitude derivatives to zero at leading order. Solving the resulting algebraic equations for each atom gives
\begin{equation}
 |\psi_{ri}\rangle\simeq-\frac{g_ra|\psi_{0i}\rangle+(\Omega_r/2)|\psi_{1i}\rangle}{\delta_r},\qquad
 |\psi_{si}\rangle\simeq-\frac{g_sa|\psi_{1i}\rangle+(\Omega_s/2)|\psi_{0i}\rangle}{\delta_s}.
 \label{eq:micro-elimination}
\end{equation}
Here the amplitudes also include the other atoms in their long-lived states. The auxiliary amplitudes still follow the evolving $|0\rangle$ and $|1\rangle$ amplitudes. Substituting these expressions back into the equations projected onto $\langle0|$ and $\langle1|$ removes the auxiliary levels and gives the effective Hamiltonian to second order in the optical couplings:
\begin{align}
 H_{\mathrm{eff}}\simeq{}&H_{\mathrm{enc}}+\Delta_0n+\epsilon\sum_iP_{1i}\nonumber\\
 &-\sum_i\left[\frac{|g_r|^2}{\delta_r}nP_{0i}+\frac{|g_s|^2}{\delta_s}nP_{1i}
 +\frac{|\Omega_r|^2}{4\delta_r}P_{1i}+\frac{|\Omega_s|^2}{4\delta_s}P_{0i}\right]\nonumber\\
 &+\sum_i\left(\beta_-a^\dagger\sigma_i^-+\beta_+a^\dagger\sigma_i^++\mathrm{H.c.}\right),
 \label{eq:micro-expanded-H}\\
 \beta_-={}&-\frac{g_r^*\Omega_r}{2\delta_r},\qquad
 \beta_+=-\frac{g_s^*\Omega_s}{2\delta_s}.
 \label{eq:micro-raman-amplitudes}
\end{align}
Here $H_{\mathrm{enc}}$ is restricted to the long-lived states. The diagonal corrections shift the cavity and spin energies. The cross terms describe two-step Raman processes: $|1,n\rangle\to|r,n\rangle\to|0,n+1\rangle$ produces $a^\dagger\sigma_i^-$, while $|0,n\rangle\to|s,n\rangle\to|1,n+1\rangle$ produces $a^\dagger\sigma_i^+$. The intermediate states are virtual, and the Hermitian conjugates describe the reverse processes.

\SMparagraph{Balanced channels and the target coupling.}
The laser frequencies set the detunings in Eq.~\eqref{eq:micro-detunings}; their amplitudes and phases set $\beta_\pm$. For a level scheme that allows balanced cavity shifts, choose these parameters so that
\begin{equation}
 \frac{|g_r|^2}{\delta_r}=\frac{|g_s|^2}{\delta_s}\equiv u,\qquad
 \beta_-=\beta_+\equiv\lambda\in\mathbb R.
 \label{eq:micro-balance}
\end{equation}
The first condition requires same-sign detunings for nonzero coupling. It makes the cavity shift independent of the spin state: since $P_{0i}+P_{1i}=I_i$, the total shift is $-Nu\,n$. The second condition combines the Raman terms into $\lambda(a+a^\dagger)S_x$, where $S_x=\sum_i(\sigma_i^++\sigma_i^-)$.

Together, these conditions also give $|\Omega_r|^2/(4\delta_r)=|\Omega_s|^2/(4\delta_s)=\lambda^2/u$ for $u\ne0$. The laser shifts are therefore equal for both spin states and contribute only the constant $-N\lambda^2/u$. The remaining local term $\epsilon\sum_iP_{1i}$ is included when choosing $H_{\mathrm{enc}}$, so that $H_{\mathrm{enc}}+\epsilon\sum_iP_{1i}=H_{\mathrm{problem}}$. Dropping the constant energy shift then gives
\begin{equation}
 \boxed{H_{\mathrm{total}}=H_{\mathrm{problem}}+\Delta a^\dagger a
 +\lambda(a+a^\dagger)S_x,\qquad \Delta=\Delta_0-Nu.}
 \label{eq:micro-target-coupling}
\end{equation}
Here $\lambda$ is the coupling per atom, with no implicit $1/\sqrt N$ factor. Both Raman processes remain in the effective model; the laser supplies the energy for $a^\dagger\sigma_i^+$. No further rotating-wave approximation is made on this interaction.

For weak coupling of the cavity to a broadband vacuum reservoir, the Born--Markov and optical rotating-wave approximations give the cavity-loss term $\kappa\mathcal D[a]$. It is unchanged by the frame rotation because $\mathcal D[e^{-i\nu t}a]=\mathcal D[a]$. Retaining this loss term at leading order gives the master equation used in the main text~\SMcite{Gorini1976,Lindblad1976},
\begin{equation}
 \dot\rho=-i[H_{\mathrm{total}},\rho]+\kappa\mathcal D[a]\rho,\qquad
 \mathcal D[a]\rho=a\rho a^\dagger-\tfrac12\{a^\dagger a,\rho\}.
 \label{eq:micro-master-equation}
\end{equation}
The cavity remains part of the dynamics here; only the two auxiliary atomic levels have been eliminated.

\SMsubsection{Graph ensembles and numerical conventions}
\label{sec:sm-graph-numerics}
The parameter scans in main-text Fig.~3 use $N=10$ random 3-regular graphs, with the same ten independent instances reused at every parameter point for each problem; $P_g(200)$ is averaged arithmetically. The size study in Fig.~4 uses 100 random connected graphs per problem and size $N=5,\ldots,12$, with ensemble-averaged degree $\overline{k}=3$, defined for $M$ instances by

\begin{equation}
 \overline{k}=\frac1M\sum_{r=1}^{M}\frac{2|\mathcal E_r|}{N}.
 \label{eq:sm-degree}
\end{equation}

The latter ensemble does not require individual graphs to be 3-regular, which would exclude odd $N$. Table~\ref{tab:sm-benchmarks} summarizes the settings. Energy and rate units are $J$ for MaxCut and $\delta$ for MIS; time units are $J^{-1}$ and $\delta^{-1}$, respectively. Both finite-coupling studies use $d_{\mathrm{photon}}=4$. Scans end at dimensionless time 200; the size study reports the median instance relaxation time defined in Sec.~\ref{sec:sm-numerics}.

\begin{table}[!htbp]
\centering
\caption{Settings reported for the main-text results. A scan varies both $\Delta$ and $\lambda$; each fixed parameter entry is dimensionless in the units shown. The finite-coupling cavity dimension is $d_{\mathrm{photon}}=4$ in both the scans and the size studies.}
\label{tab:sm-benchmarks}
\small
\setlength{\tabcolsep}{4pt}
\renewcommand{\arraystretch}{1.18}
\begin{tabular}{p{0.16\linewidth}*{4}{>{\centering\arraybackslash}p{0.17\linewidth}}}
\hline
 & MaxCut scan & MIS scan & MaxCut size study & MIS size study\\
\hline
Main-text figure & 3(a) & 3(b) & 4 & 4\\
Energy unit & $J$ & $\delta$ & $J$ & $\delta$\\
$N$ & 10 & 10 & $5,\ldots,12$ & $5,\ldots,12$\\
Graph ensemble & 3-regular & 3-regular & connected, $\langle k\rangle=3$ & connected, $\langle k\rangle=3$\\
Instances & 10 fixed & 10 fixed & 100 per size & 100 per size\\
$d_{\mathrm{photon}}$ & 4 & 4 & 4 & 4\\
$\Delta$ & scan & scan & 3.0 & 2.0\\
$\lambda$ & scan & scan & 0.25 & 0.15\\
$\kappa$ & 1.0 & 1.0 & 1.0 & 1.0\\
$J/\delta$ for MIS & --- & 2.0 & --- & 2.0\\
Reported statistic & mean $P_g(200)$ & mean $P_g(200)$ & median $t_{\mathrm{mix}}$ & median $t_{\mathrm{mix}}$\\
\hline
\end{tabular}
\end{table}

For cavity dimension $d$, use the Fock basis $q=0,\ldots,d-1$ and

\begin{equation}
 a_d=\sum_{q=1}^{d-1}\sqrt q\,|q-1\rangle\langle q| .
 \label{eq:sm-truncation}
\end{equation}

The joint Hilbert-space dimension is $D=2^Nd$, corresponding to $D^2$ complex density-matrix entries. Exact enumeration of the $2^N$ spin configurations supplies the ground-state projector and objective-value references at the simulated sizes. The integration method and photon-cutoff comparison are described in Sec.~\ref{sec:sm-numerics}.

\SMsection{Spin--photon transitions and false-vacuum escape pathways}
\label{sec:transition-network}

\SMsubsection{Composite states and allowed transitions}
Let $H_{\mathrm p}\equiv H_{\mathrm{problem}}$ and let $|\alpha\rangle$ denote a computational-basis configuration with energy $E_\alpha$. Different configuration labels may have the same energy. With $n=a^\dagger a$ and $H_0=H_{\mathrm p}+\Delta n$, the uncoupled composite states are $|\alpha,q\rangle=|\alpha\rangle\otimes|q\rangle$, with energy $E_\alpha+q\Delta$. Define
\begin{equation}
 M_{\beta\alpha}=\langle\beta|S_x|\alpha\rangle,
 \qquad S_x=\sum_{i=1}^{N}\sigma_i^x .
 \label{eq:theory-matrix-element}
\end{equation}
For the unweighted collective coupling used here, $M_{\beta\alpha}=1$ when the two configurations differ by one spin flip and is zero otherwise. This statement refers to the computational basis; matrix elements in a symmetry-adapted basis need not have this form. The interaction $H_{\mathrm{coupling}}=\lambda(a+a^\dagger)S_x$ has matrix elements
\begin{align}
 \langle\beta,q+1|H_{\mathrm{coupling}}|\alpha,q\rangle
 &=\lambda\sqrt{q+1}\,M_{\beta\alpha},\\
 \langle\beta,q-1|H_{\mathrm{coupling}}|\alpha,q\rangle
 &=\lambda\sqrt q\,M_{\beta\alpha}.
 \label{eq:theory-composite-edges}
\end{align}
The corresponding changes in bare composite energy are
\begin{equation}
 \delta^{(+)}_{\beta\alpha}=E_\beta-E_\alpha+\Delta,
 \qquad
 \delta^{(-)}_{\beta\alpha}=E_\beta-E_\alpha-\Delta.
 \label{eq:theory-edge-detunings}
\end{equation}
Thus a downward spin transition with $E_\alpha-E_\beta=\Delta>0$ can resonantly create a photon. An upward transition of the same magnitude can resonantly absorb a photon, but its matrix element vanishes at $q=0$. The cavity produces photon-number replicas of the spin energy landscape, linked by these coherent transitions. Their matrix elements, interference, detuning, and leakage jointly determine population transfer; the existence of an allowed path alone does not determine its transfer rate.

Photon leakage is described by the jump operator $L=\sqrt{\kappa}\,a$, with $L|\alpha,q\rangle=\sqrt{\kappa q}\,|\alpha,q-1\rangle$ for $q>0$. It removes a photon without flipping a spin and lowers the bare composite energy $H_0$ by $\Delta$. Since $[H_{\mathrm p},a]=0$, its dissipative contribution to the ensemble-averaged spin energy vanishes, $\mathcal D^\dagger[a](H_{\mathrm p})=0$. Spin relaxation therefore relies on coherent spin--photon exchange acting together with leakage. At the finite-size benchmark settings $\Delta=3J$ for MaxCut and $\Delta=2\delta$ for MIS with $J=2\delta$, no single-spin-flip transition is exactly resonant with one photon. Their dynamics must consequently be assessed using the full Hamiltonian with finite coupling and linewidth; monotonic spin cooling and an instance-independent escape rate do not follow from the allowed transitions alone.

\SMsubsection{Configuration space interpretation of false vacuum decay}
In the optimization landscape, a strict one-spin-flip local minimum is a configuration for which every single spin flip raises $H_{\mathrm p}$. A nonglobal local minimum provides the operational false-vacuum state, while the ground-state manifold supplies the true-vacuum reference. This is an analogy with metastable decay in configuration space, rather than a claim of real-space bubble nucleation or a field-theoretic instanton description~\SMcite{Coleman1977,CallanColeman1977}.

For an ideal resonant realization of the illustrative path in main-text Fig.~2, take a photon-occupied state $|\ell,n\rangle$ with $n>0$, a bridge configuration $|b\rangle$, a distinct configuration $|\ell'\rangle$ outside the initial basin with $E_{\ell'}=E_\ell$, and a lower-energy configuration $|d\rangle$. Each adjacent pair along the proposed path must differ by one spin flip. The two stages are
\begin{align}
 |\ell,n\rangle&\leftrightarrow|b,n-1\rangle
 \leftrightarrow|\ell',n\rangle,
 &\delta_{\mathrm b}&=E_b-E_\ell-\Delta,\\
 |\ell',n\rangle&\leftrightarrow|d,n+1\rangle
 \xrightarrow{\,\kappa(n+1)\,}|d,n\rangle,
 &\delta_d&=E_{\ell'}-E_d-\Delta.
 \label{eq:theory-fvd-path}
\end{align}
The first stage is resonant for $\delta_{\mathrm b}=0$ and has coherent amplitudes proportional to $\lambda\sqrt n$; the second coherent step is resonant for $\delta_d=0$ and has amplitude proportional to $\lambda\sqrt{n+1}$. The final arrow denotes a transition rate, not an amplitude. Photon leakage reduces the cavity occupation from $n+1$ to $n$ while retaining the lower-energy spin configuration $|d\rangle$. This dissipative step interrupts the direct coherent reversal of the preceding photon-creation step; return through other transitions remains possible.

The path is conditional on occupation of a finite-photon sector. Because the protocol starts in cavity vacuum, the full dynamics must first generate such occupation; the absorption step cannot operate directly from $n=0$. Furthermore, resonant three-state chains may support dark superpositions, so the existence of the displayed path alone does not establish a nonzero decay rate for every initial state. Near-resonant transfer, competing paths, and leakage must be assessed in the complete dynamics. The present theoretical construction identifies allowed escape and relaxation channels. Quantitative basin survival, doorway occupation, and decay-rate measurements require basin-resolved simulations and are not asserted here.

\SMsection{Weak-coupling dynamics and stationary states}
\label{sec:weak-coupling-steady-state}
We derive the operator $\mathcal K_2$ in main-text Eq.~(2), which constrains the stationary spin state at weak spin--photon coupling. We first identify the uncoupled stationary subspace, then find the first-order response outside it and obtain the second-order condition within it. We keep the problem instance, $\Delta$, and $\kappa>0$ fixed and first work at a finite photon cutoff.

\SMsubsection{Unperturbed stationary subspace}
Separate the Liouvillian into the uncoupled evolution and the spin--photon interaction:
\begin{equation}
 \begin{aligned}
 \mathcal L(\lambda)&=\mathcal L_0+\lambda\mathcal V,\\
 \mathcal L_0X&=-i[H_{\mathrm p}+\Delta n,X]+\kappa\mathcal D[a]X,\\
 \mathcal VX&=-i[(a+a^\dagger)S_x,X].
 \end{aligned}
 \label{eq:theory-L0-V}
\end{equation}
Here $X$ is a joint spin--cavity operator, $n=a^\dagger a$, and $\mathcal D[a]X=aXa^\dagger-\{n,X\}/2$. Write $H_{\mathrm p}=\sum_E E\Pi_E$, where $\Pi_E$ projects onto all spin states with energy $E$. At $\lambda=0$, cavity loss drives the photons to the vacuum $\tau_0=|0\rangle\langle0|$. The spins evolve only under $H_{\mathrm p}$, so a spin operator $Y$ is stationary precisely when $[H_{\mathrm p},Y]=0$. Therefore,
\begin{equation}
 \ker\mathcal L_0
 =\{Y\otimes\tau_0:\ [H_{\mathrm p},Y]=0\}.
 \label{eq:theory-L0-kernel}
\end{equation}
This kernel contains all stationary operators. Since it generally includes many physical states, $\mathcal L_0\rho_0=0$ alone cannot select the weak-coupling steady state.

To separate stationary and nonstationary components, define
\begin{equation}
 \begin{aligned}
 \mathcal E_{\mathrm p}(Y)&=\sum_E\Pi_EY\Pi_E,\\
 \mathcal PX&=\mathcal E_{\mathrm p}\big(\operatorname{Tr}_{\mathrm{cav}}X\big)
 \otimes\tau_0,\qquad \mathcal Q=\mathcal I-\mathcal P.
 \end{aligned}
 \label{eq:theory-PQ}
\end{equation}
The map $\mathcal E_{\mathrm p}$ removes coherences between different spin energies but retains those within a degenerate energy block. The projection $\mathcal P$ first traces out the cavity, applies this map, and restores the cavity vacuum. Thus it extracts the stationary component, including populations that reach the vacuum through photon loss. The complementary projection $\mathcal Q$ contains the remaining oscillating or decaying components. These maps obey $\mathcal P^2=\mathcal P$ and $\mathcal L_0\mathcal P=\mathcal P\mathcal L_0=0$. 

\SMsubsection{Derivation of the second-order spin generator}
The next step requires solving equations of the form $\mathcal L_0X=Z$, where $Z$ is a known source term. The full $\mathcal L_0$ has no inverse because it sends every stationary component to zero. It is invertible, however, on $\operatorname{Ran}\mathcal Q$, the space of operators satisfying $\mathcal QX=X$. We therefore define the reduced inverse
\begin{equation}
 \mathcal R=\mathcal Q
 \left(\mathcal L_0\big|_{\operatorname{Ran}\mathcal Q}\right)^{-1}\mathcal Q,
 \qquad \mathcal L_0\mathcal R=\mathcal R\mathcal L_0=\mathcal Q.
 \label{eq:theory-reduced-inverse}
\end{equation}
Applying $\mathcal P$ to $\mathcal L_0X=Z$ gives $0=\mathcal PZ$: the source cannot contain a stationary component. Conversely, when $\mathcal PZ=0$, the identity $\mathcal L_0\mathcal RZ=\mathcal QZ=Z$ shows that $\mathcal RZ$ solves the equation. The general solution is $X=\mathcal RZ+\mathcal PX$, with an arbitrary stationary part.

Assume a regular expansion $\rho_{\mathrm{ss}}(\lambda)=\rho_0+\lambda\rho_1+\lambda^2\rho_2+\cdots$. Substituting it into $\mathcal L(\lambda)\rho_{\mathrm{ss}}(\lambda)=0$ and matching powers of $\lambda$ gives
\begin{equation}
 \mathcal L_0\rho_0=0,\qquad
 \mathcal L_0\rho_1=-\mathcal V\rho_0,\qquad
 \mathcal L_0\rho_2=-\mathcal V\rho_1.
 \label{eq:theory-stationary-hierarchy}
\end{equation}
The first equation gives $\rho_0=\sigma_0\otimes\tau_0$, with $[H_{\mathrm p},\sigma_0]=0$. The second determines the nonstationary part of $\rho_1$. Since the vacuum has zero field amplitude, $\operatorname{Tr}[(a+a^\dagger)\tau_0]=0$, one interaction has no component inside the stationary subspace. Hence
\begin{equation}
 \mathcal P\mathcal V\mathcal P=0,\qquad
 \rho_1=\mathcal P\rho_1-\mathcal R\mathcal V\rho_0.
 \label{eq:theory-general-rho1}
\end{equation}
The first-order solvability condition is therefore automatic and does not select $\sigma_0$; it does not mean that $\rho_1$ vanishes. To obtain the next condition, apply $\mathcal P$ to the equation for $\rho_2$ and substitute the expression for $\rho_1$:
\begin{equation}
 0=\mathcal P\mathcal V\rho_1
 =\underbrace{\mathcal P\mathcal V\mathcal P\rho_1}_{=0}
 -\mathcal P\mathcal V\mathcal R\mathcal V\rho_0.
 \label{eq:theory-second-order-condition}
\end{equation}
Since $\mathcal P\mathcal V\mathcal P=0$, the unknown $\mathcal P\rho_1$ drops out, leaving a condition on $\rho_0$ that defines the second-order spin generator~\SMcite{Kessler2012}:
\begin{equation}
 (\mathcal K_2\sigma)\otimes\tau_0
 =-\mathcal P\mathcal V\mathcal R\mathcal V(\sigma\otimes\tau_0),
 \qquad [H_{\mathrm p},\sigma]=0.
 \label{eq:theory-K2-definition}
\end{equation}
Reading the operators from right to left, the first $\mathcal V$ takes the state out of the stationary subspace, $\mathcal R$ gives the uncoupled response, and the second $\mathcal V$ followed by $\mathcal P$ returns its contribution to that subspace. There are two interaction factors, so the physical slow generator is $\lambda^2\mathcal K_2$ to leading order. This factor sets the timescale, whereas stationary states must satisfy $\mathcal K_2\sigma_0=0$:
\begin{equation}
 \rho_{\mathrm{ss}}(\lambda)=\sigma_0\otimes\tau_0+O(\lambda),
 \qquad \mathcal K_2\sigma_0=0,
 \qquad \sigma_0\succeq0,\quad\operatorname{Tr}\sigma_0=1.
 \label{eq:theory-leading-steady-state}
\end{equation}
Here $\sigma_0$ is the main-text $\rho_s^{(0)}$; positivity and unit trace make it a physical spin state. Its zeroth-order success probability is
\begin{equation}
 P_g^{(0)}=\operatorname{Tr}(\Pi_g\sigma_0),
 \label{eq:theory-leading-ground-probability}
\end{equation}
where $\Pi_g$ projects onto all optimal spin configurations. These are necessary leading-order conditions. If $\mathcal K_2$ has several stationary states, conserved quantities and possibly higher-order terms are needed to determine the selected state. The stationary weak-coupling limit takes the long-time limit first, so $\sigma_0$ generally differs from the uncoupled projection of the initial state.

At fixed finite instance and photon cutoff, $\mathcal K_2$ also governs the projected dynamics as $\lambda\to0$ with $\tau=\lambda^2t$ fixed on bounded intervals. Since $\Pi_g$ is energy diagonal, its slow dynamics can be evaluated from the energy-projected preparation $\mathcal E_{\mathrm p}(|+\rangle\langle+|^{\otimes N})$, retaining coherence within each degenerate block.

\SMsubsection{Ground-state probabilities in the effective model}
We obtain the numerical results by integrating $\partial_\tau\sigma=\mathcal K_2\sigma$ with the quantum-trajectory, or Monte Carlo wave-function (MCWF), method~\SMcite{Dalibard1992,Molmer1993}. Each graph is simulated with an ensemble of stochastic pure-state trajectories, and the ground-state probability is estimated from their average. We use the slow time $\tau=\lambda^2t$ in the dimensionless units of Table~\ref{tab:sm-benchmarks}, with
\begin{equation}
 \sigma_{\mathrm{in}}=\mathcal E_{\mathrm p}\big(|+\rangle\langle+|^{\otimes N}\big),
 \qquad P_{g,\mathrm{eff}}(\tau)
 =\operatorname{Tr}\!\left[\Pi_g e^{\tau\mathcal K_2}\sigma_{\mathrm{in}}\right].
 \label{eq:theory-finite-time-probability}
\end{equation}

For each graph, we compute the distinct energies of $H_{\mathrm p}$ and group the single-spin-flip matrix elements by their energy difference $\omega=E-E'$. The jump operators and coherent energy-shift operator used to implement Eq.~\eqref{eq:theory-K2-definition} are
\begin{equation}
 \begin{aligned}
 A_\omega&=\sum_{E-E'=\omega}\Pi_{E'}S_x\Pi_E,
 &L_\omega&=\sqrt{\frac{\kappa}{(\omega-\Delta)^2+(\kappa/2)^2}}\,A_\omega,\\
 H_{\mathrm{LS}}&=\sum_\omega
 \frac{\omega-\Delta}{(\omega-\Delta)^2+(\kappa/2)^2}
 A_\omega^\dagger A_\omega.
 \end{aligned}
 \label{eq:theory-mcwf-coefficients}
\end{equation}
For an energy-block-diagonal state, these operators give $\mathcal K_2\sigma=-i[H_{\mathrm{LS}},\sigma]+\sum_\omega\mathcal D[L_\omega]\sigma$. All allowed energy differences are included. To sample $\sigma_{\mathrm{in}}$, each trajectory starts by selecting an energy $E$ with probability $p_E=\|\Pi_E|\psi_+\rangle\|^2$, where $|\psi_+\rangle=|+\rangle^{\otimes N}$, and setting $|\psi(0)\rangle=\Pi_E|\psi_+\rangle/\sqrt{p_E}$ for $p_E>0$.

Between jumps, we propagate the state using
\begin{equation}
 |\widetilde\psi(\tau+\delta\tau)\rangle
 =e^{-iH_{\mathrm{nh}}\delta\tau}|\psi(\tau)\rangle,
 \qquad H_{\mathrm{nh}}=H_{\mathrm{LS}}
 -\frac{i}{2}\sum_\omega L_\omega^\dagger L_\omega,
 \label{eq:theory-mcwf-no-jump}
\end{equation}
and normalize it after propagation. The matrix-exponential action is evaluated with a sparse numerical routine. Jump times are sampled by continuous-time Poisson thinning: an upper bound $R$ valid throughout the no-jump evolution in the current energy block sets exponentially distributed candidate waiting times. At each candidate time, we calculate $r_\omega=\|L_\omega|\psi\rangle\|^2$ from the propagated state and accept a jump with probability $r/R$, where $r=\sum_\omega r_\omega$. An accepted jump selects channel $\omega$ with probability $r_\omega/r$ and updates the state to $L_\omega|\psi\rangle/\sqrt{r_\omega}$; otherwise, the propagated state is retained. If the next candidate lies beyond the final time, we propagate only to that final time.

For each graph we run $K=512$ independent trajectories and calculate
\begin{equation}
 \widehat P_g(\tau)=\frac1K\sum_{\ell=1}^K
 \langle\psi_\ell(\tau)|\Pi_g|\psi_\ell(\tau)\rangle.
 \label{eq:theory-mcwf-estimators}
\end{equation}
The probability estimate is accumulated directly from the trajectories without storing a density matrix.

For the effective-model simulations, we use $J=1$, $\Delta=3$, $\kappa=1$, and a final time $\tau=250$ for MaxCut. For MIS, we use $\delta=1$, $J=2$, $\Delta=2$, $\kappa=1$, and $\tau=300$. The coupling factor $\lambda^2$ is absorbed into the slow time $\tau=\lambda^2t$. 

For each problem, the effective-model dataset contains 1410 labeled samples of connected Erd\H{o}s--R\'enyi graphs with $N=5,\ldots,18$. The sample counts, listed in increasing order of $N$, are $(40,50,60,70,80,90,100,110,120,130,140,140,140,140)$. Candidate graphs are generated by connecting each vertex pair independently with probability $p=3/(N-1)$, and only connected graphs are retained. We do not require all vertices to have the same degree. Repeated or isomorphic samples are retained.

For comparison, the finite-coupling solver uses $\lambda=0.25$ for MaxCut and $\lambda=0.15$ for MIS. Its dataset contains 800 records per problem over $N=5,\ldots,12$, with $P_g$ evaluated at the onset of the detected plateau region. Figure~\ref{fig:sm-zero-order-means} shows the mean ground-state probabilities and their standard errors for both methods.

\begin{figure}[!t]
\centering
\begin{tikzpicture}
\definecolor{s2blue}{RGB}{0,114,178}
\definecolor{s2orange}{RGB}{213,94,0}
\begin{groupplot}[
  group style={group size=2 by 1,horizontal sep=0.045\textwidth},
  width=0.445\textwidth,height=0.29\textwidth,
  /tikz/font=\small,
  xmin=4.5,xmax=18.5,ymin=0,ymax=1,
  xtick={5,7,9,11,13,15,18},
  ytick={0,0.2,0.4,0.6,0.8,1},
  scaled ticks=false,
  xlabel={Number of spins, $N$},
  ylabel={Mean ground-state probability, $\overline{P}_g$},
  tick align=inside,
  tick style={line width=0.6pt},
  axis line style={line width=0.6pt},
  ymajorgrids=true,
  grid style={draw=gray!25,line width=0.3pt},
  title style={at={(0,1.035)},anchor=south west},
  legend columns=2,
  legend style={draw=none,font=\small,cells={anchor=west},
    /tikz/every even column/.append style={column sep=1.2em}}
]
\nextgroupplot[title={(a) MaxCut},legend to name=s2groundlegend]
\addplot+[
  color=s2blue,
  mark=o,
  mark size=2.1pt,
  mark options={solid,fill=white,line width=0.8pt},
  line width=0.75pt,
  error bars/.cd,
  y dir=both,y explicit,
  error bar style={line width=0.6pt},
  error mark options={rotate=90,mark size=1.7pt,line width=0.6pt},
  /pgfplots/.cd
] table[x=N,y=mean,y error=SE] {sm_maxcut_effective.dat};
\addlegendentry{Second-order effective model}
\addplot+[
  color=s2orange,
  dashed,
  mark=square,
  mark size=2.1pt,
  mark options={solid,fill=white,line width=0.8pt},
  line width=0.75pt,
  error bars/.cd,
  y dir=both,y explicit,
  error bar style={line width=0.6pt},
  error mark options={rotate=90,mark size=1.7pt,line width=0.6pt},
  /pgfplots/.cd
] table[x=N,y=mean,y error=SE] {sm_maxcut_solver.dat};
\addlegendentry{Finite-coupling solver}
\nextgroupplot[title={(b) MIS},ylabel={},yticklabels={}]
\addplot+[
  color=s2blue,
  mark=o,
  mark size=2.1pt,
  mark options={solid,fill=white,line width=0.8pt},
  line width=0.75pt,
  error bars/.cd,
  y dir=both,y explicit,
  error bar style={line width=0.6pt},
  error mark options={rotate=90,mark size=1.7pt,line width=0.6pt},
  /pgfplots/.cd,forget plot
] table[x=N,y=mean,y error=SE] {sm_mis_effective.dat};
\addplot+[
  color=s2orange,
  dashed,
  mark=square,
  mark size=2.1pt,
  mark options={solid,fill=white,line width=0.8pt},
  line width=0.75pt,
  error bars/.cd,
  y dir=both,y explicit,
  error bar style={line width=0.6pt},
  error mark options={rotate=90,mark size=1.7pt,line width=0.6pt},
  /pgfplots/.cd,forget plot
] table[x=N,y=mean,y error=SE] {sm_mis_solver.dat};
\end{groupplot}
\path (group c1r1.north west) -- (group c2r1.north east)
  node[midway,above=9mm] {\pgfplotslegendfromname{s2groundlegend}};
\end{tikzpicture}
\caption{Ensemble-averaged ground-manifold probability for (a) MaxCut and (b) MIS. Circles show the effective-model results at $\tau=250$ and $300$, respectively; squares show the finite-coupling solver at the plateau-region onset, with $\lambda=0.25$ and $0.15$. Error bars show one standard error of the mean across graph samples. Lines connect the mean values to show the trend. The effective-model data cover $N=5,\ldots,18$, while solver records are available for $N=5,\ldots,12$.}
\label{fig:sm-zero-order-means}
\end{figure}
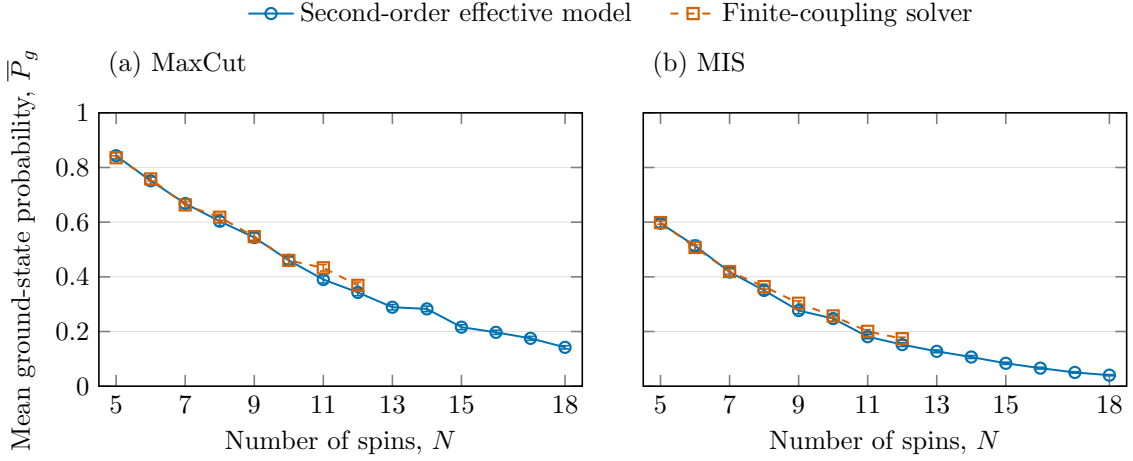

For each size $N$, we average the ground-state probabilities over the $m_N$ graph samples to obtain $\overline P_N$. The standard error of this mean is $\mathrm{SE}_N=s_N/\sqrt{m_N}$, where $s_N$ is the sample standard deviation. We fit these mean probabilities directly to an exponential function by minimizing
\begin{equation}
 \overline P_N\simeq A\,2^{bN},\qquad
 \chi^2(A,b)=\sum_N
 \left[\frac{\overline P_N-A\,2^{bN}}{\mathrm{SE}_N}\right]^2.
 \label{eq:theory-numerical-weighted-fit}
\end{equation}
Here $A$ sets the overall scale, and $b$ describes how the probability changes with $N$. The weights $1/\mathrm{SE}_N^2$ give more weight to means with smaller errors. Table~\ref{tab:sm-zero-order-fits} lists the fitted parameters and $R^2$. The reported $R^2$ is calculated without weights, using the original probability values.

To estimate the uncertainty in $b$, we repeatedly draw samples from the existing graphs at each $N$. Each sample has the same number of graphs as the original, but a graph may be selected more than once. We recalculate the mean probabilities and fit the curve again using the original weights. After 4000 such repetitions, we use the resulting $b$ values to estimate the 95\% confidence interval (bootstrap method).
\begin{table}[!t]
\centering
\caption{Final fits $\overline P_N=A\,2^{bN}$, with weights $1/\mathrm{SE}_N^2$. The stated $R^2$ is evaluated without weights. The confidence interval (CI) is the bootstrap 95\% interval for $b$ and does not bound finite-time bias.}
\label{tab:sm-zero-order-fits}
\small
\setlength{\tabcolsep}{4pt}
\renewcommand{\arraystretch}{1.12}
\begin{tabular}{llrrrrl}
\toprule
Problem & Method & $N$ range & $A$ & $b$ & $R^2$ & 95\% CI for $b$\\
\midrule
MaxCut & Effective & 5--18 & 1.67556 & $-0.19170$ & 0.99469 & $[-0.19664,-0.18677]$\\
MaxCut & Solver & 5--12 & 1.48078 & $-0.16318$ & 0.99387 & $[-0.17079,-0.15551]$\\
MIS & Effective & 5--18 & 1.71223 & $-0.29289$ & 0.99647 & $[-0.29984,-0.28576]$\\
MIS & Solver & 5--12 & 1.44563 & $-0.25339$ & 0.99811 & $[-0.26217,-0.24457]$\\
\bottomrule
\end{tabular}
\end{table}

\SMsection{Numerical methods and performance estimates}
\label{sec:sm-numerics}

\SMsubsection{Numerical integration and photon cutoff}
The finite-coupling solver evolves the full spin-cavity density matrix using a fourth-order Runge-Kutta (RK4) method. With $H=H_{\mathrm{problem}}+\Delta a^\dagger a+\lambda(a+a^\dagger)S_x$, the right-hand side evaluated numerically is
\begin{equation}
 \dot\rho=\mathcal F(\rho)
 =-i[H,\rho]+\kappa\left(a\rho a^\dagger-\frac12\{a^\dagger a,\rho\}\right).
 \label{eq:sm-rk4-rhs}
\end{equation}
We work in the basis $|z,q\rangle$, where $z$ labels a spin configuration and $q=0,\ldots,d_{\mathrm{photon}}-1$, with $d_{\mathrm{photon}}=4$ for the main-text benchmarks. We initialize $\rho(0)=(|+\rangle\langle+|)^{\otimes N}\otimes|0\rangle\langle0|$. Writing $\rho_m$ for the numerical state at $t_m=mh$, one step is
\begin{equation}
 \begin{aligned}
 k_1&=\mathcal F(\rho_m),\qquad
 k_2=\mathcal F\!\left(\rho_m+\frac h2 k_1\right),\\
 k_3&=\mathcal F\!\left(\rho_m+\frac h2 k_2\right),\qquad
 k_4=\mathcal F(\rho_m+h k_3),\\
 \widetilde\rho_{m+1}&=\rho_m+\frac h6(k_1+2k_2+2k_3+k_4).
 \end{aligned}
 \label{eq:sm-rk4-update}
\end{equation}
Before the next step, $\widetilde\rho_{m+1}$ is symmetrized as $(\widetilde\rho_{m+1}+\widetilde\rho_{m+1}^\dagger)/2$ and normalized by its trace to obtain $\rho_{m+1}$. The size study uses $h=0.05$, in the time units specified in Sec.~\ref{sec:sm-graph-numerics}.

At each completed step, $P_g(t)$ is obtained by summing the diagonal populations over all photon states and all ground-state spin configurations, identified by exact enumeration for each graph. Thus, in the size study, the recording interval equals the integration step. The parameter scans are evolved to $t=200$ and averaged over the fixed graph samples. The main-text benchmark Hamiltonian and loss parameters are listed in Table~\ref{tab:sm-benchmarks}.

To motivate our choice of $d_{\mathrm{photon}}=4$, Fig.~\ref{fig:sm-photon-cutoff} compares $P_g(t)$ for $d_{\mathrm{photon}}=2,4,6,8$ using one $N=10$ MaxCut graph and the parameters stated in the caption. This is a separate cutoff test from the main-text size study. While $d_{\mathrm{photon}}=2$ gives a noticeably different curve, the results for $d_{\mathrm{photon}}=4,6,8$ nearly coincide after the initial transient. Over $10\leq t\leq200$, increasing the dimension from 4 to 6 or 8 changes $P_g(t)$ by less than $2.2\times10^{-3}$. At $t=200$, the three probabilities are $0.4055$, $0.4042$, and $0.4043$, respectively. The inset shows the small differences at early times. This comparison supports $d_{\mathrm{photon}}=4$ as a practical balance between numerical accuracy and computational cost for the tested instance and parameters.

\begin{figure}[!htbp]
\centering
\includegraphics[width=0.84\linewidth]{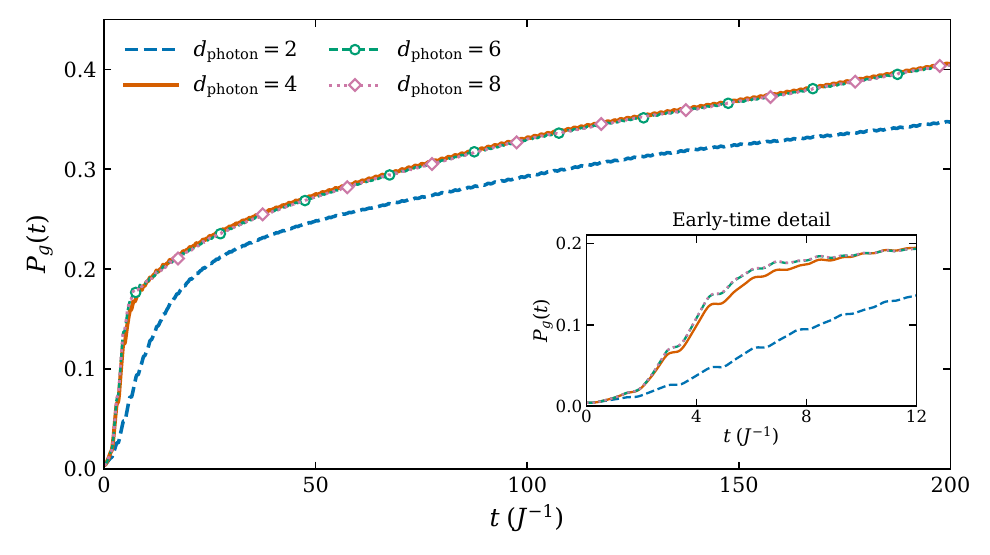}
\caption{Dependence of the ground-state probability on the cavity dimension. The four curves use the same ten-node 3-regular MaxCut graph, with $\Delta/J=2$, $\lambda/J=0.15$, and $\kappa/J=1$. The initial state is $(|+\rangle\langle+|)^{\otimes N}\otimes|0\rangle\langle0|$, and the RK4 time step is $h=0.05J^{-1}$. The inset enlarges the initial transient.}
\label{fig:sm-photon-cutoff}
\end{figure}

\SMsubsection{Mixing time and success probability}
For each graph instance $r$, we record $P_g^{(r)}(t)$ every $\delta t=0.05$. At each candidate starting time $T_0$ with $P_g^{(r)}(T_0)>0$, we check the next 1000 recorded values, covering 50 time units. The mixing time is the earliest $T_0$ for which all these values differ from $P_g^{(r)}(T_0)$ by at most $1\%$ of that reference value:
\begin{equation}
 t_{\mathrm{mix}}^{(r)}=\inf\left\{T_0:
 \max_{1\leq j\leq1000}
 \frac{|P_g^{(r)}(T_0+j\delta t)-P_g^{(r)}(T_0)|}
 {P_g^{(r)}(T_0)}\leq0.01\right\},
 \qquad \delta t=0.05.
 \label{eq:sm-mixing}
\end{equation}
Here $\inf$ selects the earliest sampled starting time $T_0$ that satisfies the above criterion.

For each problem and size $N$, we determine the mixing times for 100 graph instances and sort them in ascending order. The median mixing time shown in main-text Fig.~4 is the average of the 50th and 51st values in this ordered list. Its error bars estimate bootstrap uncertainty in the median, and the MaxCut fit band is a 95\% bootstrap confidence interval~\SMcite{Efron1979}. 

For the main-text size study, the mean preparation probability is evaluated at each instance's own plateau onset,
\begin{equation}
 \overline{P_g^{\mathrm{mix}}}(N)=\frac1M\sum_{r=1}^{M}
 P_g^{(r)}\!\left(t_{\mathrm{mix}}^{(r)}\right),\qquad M=100,
 \label{eq:sm-mean-mixing}
\end{equation}
rather than at the ensemble median time. Figure~\ref{fig:sm-zero-order-means} compares the mean ground-state probabilities from the effective model and the full solver, while Table~\ref{tab:sm-zero-order-fits} lists the corresponding fit results. The graph ensembles, sample counts, and evolution settings are specified in Secs.~\ref{sec:sm-graph-numerics} and~\ref{sec:weak-coupling-steady-state}.

\SMsubsection{Qualitative estimate of time complexity}
\label{sec:sm-time-complexity}
We estimate the time complexity by multiplying the duration of one solver run by the number of independent repetitions needed to reach a fixed target success probability $P_{\mathrm{tar}}$. For this qualitative estimate, we use the median mixing time as the run duration $t_{\mathrm{run}}(N)$ and the mean preparation probability in Eq.~\eqref{eq:sm-mean-mixing} as the single-run success probability $p(N)$. After $R$ independent runs, the probability of obtaining an optimal solution at least once is $1-[1-p(N)]^R$. Therefore,
\begin{equation}
\begin{aligned}
 R(N)&=\left\lceil\frac{\ln(1-P_{\mathrm{tar}})}{\ln[1-p(N)]}\right\rceil,\\
 T(N)&=R(N)t_{\mathrm{run}}(N)\propto\frac{t_{\mathrm{run}}(N)}{p(N)},
\end{aligned}
\label{eq:sm-repetition-time}
\end{equation}
where the proportionality holds for small $p(N)$ at fixed $P_{\mathrm{tar}}$.

Over the studied sizes $N=5,\ldots,12$, the median mixing time in main-text Fig.~4 grows approximately linearly with $N$ for MaxCut and remains roughly constant for MIS. The finite-coupling solver fits in Table~\ref{tab:sm-zero-order-fits} give $p(N)\propto2^{-\alpha N}$, with $\alpha\approx0.16$ for MaxCut and $\alpha\approx0.25$ for MIS. Thus, the repetition count scales as $2^{0.16N}$ for MaxCut, with each run taking a time proportional to $N$, and as $2^{0.25N}$ for MIS, with each run taking an approximately constant time. If these observed trends persist at larger sizes, Eq.~\eqref{eq:sm-repetition-time} gives the qualitative time-complexity estimates
\begin{equation}
 T_{\mathrm{MaxCut}}(N)=\mathcal O\!\left(N2^{0.16N}\right),
 \qquad
 T_{\mathrm{MIS}}(N)=\mathcal O\!\left(2^{0.25N}\right).
 \label{eq:sm-time-scaling}
\end{equation}
These estimates count only the solver's total evolution time; state preparation, readout, and control costs are outside the present analysis.

\clearpage
\phantomsection
\addcontentsline{smtoc}{section}{References}
\begin{SMbibliography}
\SMbibitem{Dimer2007}F. Dimer, B. Estienne, A. S. Parkins, and H. J. Carmichael, Proposed realization of the Dicke-model quantum phase transition in an optical cavity QED system, Phys. Rev. A \textbf{75}, 013804 (2007), \href{https://doi.org/10.1103/PhysRevA.75.013804}{doi:10.1103/PhysRevA.75.013804}.
\SMbibitem{ReiterSorensen2012}F. Reiter and A. S. S{\o}rensen, Effective operator formalism for open quantum systems, Phys. Rev. A \textbf{85}, 032111 (2012), \href{https://doi.org/10.1103/PhysRevA.85.032111}{doi:10.1103/PhysRevA.85.032111}.
\SMbibitem{Gorini1976}V. Gorini, A. Kossakowski, and E. C. G. Sudarshan, Completely positive dynamical semigroups of $N$-level systems, J. Math. Phys. \textbf{17}, 821--825 (1976).
\SMbibitem{Lindblad1976}G. Lindblad, On the generators of quantum dynamical semigroups, Commun. Math. Phys. \textbf{48}, 119--130 (1976).
\SMbibitem{Coleman1977}S. Coleman, ``Fate of the false vacuum: Semiclassical theory,''
\emph{Phys. Rev. D} \textbf{15}, 2929--2936 (1977),
\href{https://doi.org/10.1103/PhysRevD.15.2929}{doi:10.1103/PhysRevD.15.2929}.
\SMbibitem{CallanColeman1977}C. G. Callan, Jr. and S. Coleman, ``Fate of the false vacuum. II. First quantum corrections,''
\emph{Phys. Rev. D} \textbf{16}, 1762--1768 (1977),
\href{https://doi.org/10.1103/PhysRevD.16.1762}{doi:10.1103/PhysRevD.16.1762}.
\SMbibitem{Kessler2012}E. M. Kessler, ``Generalized Schrieffer-Wolff formalism for dissipative systems,''
\emph{Phys. Rev. A} \textbf{86}, 012126 (2012),
\href{https://doi.org/10.1103/PhysRevA.86.012126}{doi:10.1103/PhysRevA.86.012126}.
\SMbibitem{Dalibard1992}J. Dalibard, Y. Castin, and K. M{\o}lmer, Wave-function approach to dissipative processes in quantum optics, Phys. Rev. Lett. \textbf{68}, 580--583 (1992), \href{https://doi.org/10.1103/PhysRevLett.68.580}{doi:10.1103/PhysRevLett.68.580}.
\SMbibitem{Molmer1993}K. M{\o}lmer, Y. Castin, and J. Dalibard, Monte Carlo wave-function method in quantum optics, J. Opt. Soc. Am. B \textbf{10}, 524--538 (1993), \href{https://doi.org/10.1364/JOSAB.10.000524}{doi:10.1364/JOSAB.10.000524}.
\SMbibitem{Efron1979}B. Efron, Bootstrap methods: Another look at the jackknife, Ann. Stat. \textbf{7}, 1--26 (1979).
\end{SMbibliography}

\clearpage
\endgroup
\end{document}